\RequirePackage{fix-cm}
\RequirePackage{amsmath}
\documentclass{svjour3} 

\smartqed  

\usepackage{cite}
\usepackage{graphicx}
\usepackage{amssymb}
\usepackage{bm} 

\usepackage{booktabs}
\usepackage{epstopdf}
\usepackage{pgf}
\usepackage[caption=false]{subfig}
\usepackage[section]{placeins} 

\usepackage[hidelinks]{hyperref} 
\hypersetup{
    colorlinks,
    citecolor=black,
    filecolor=black,
    linkcolor=black,
    urlcolor=black
}

\usepackage{todonotes}

\graphicspath{ {./fig-golf/} }
\journalname{Sports Engineering}

\begin{document}

\title{High-Order Computational Fluid Dynamics Simulations of a Spinning Golf Ball}


\titlerunning{High-Order CFD of a Spinning Golf Ball} 

\author{Jacob Crabill \and Freddie Witherden \and Antony Jameson}


\institute{J. Crabill \at Stanford University \\ \email{jcrabill@stanford.edu}  \and  F.D. Witherden \at Stanford University\\ \email{fdw@stanford.edu}  \and  A. Jameson \at Stanford University \\ \email{ajameson@stanford.edu}}

\date{Received: date / Accepted: date}

\maketitle

\begin{abstract}
This paper presents the first high-order computational fluid dynamics (CFD) simulations of static and spinning golf balls at realistic flow conditions.  The present results are shown to capture the complex fluid dynamics inside the dimples which lead to drag reduction versus a smooth sphere, and compare well to previous experimental and computational studies.  The high--order Flux Reconstruction method has been paired with the Artificial Boundary overset method to enable simplified mesh generation and grid motion.  The compressible Navier--Stokes equations are modeled using a scale--resolving Large Eddy Simulation (LES) approach with no sub--grid models.  The codes implementing these methods have been implemented for NVIDIA Graphical Processing Units (GPUs), enabling large speedups over traditional computer hardware.  The new method allows for the simulation of golf balls, and other objects at similar moderate Reynolds numbers, to be simulated in a matter of days on large computing clusters.  The use of CFD for the design of objects such as golf balls and other sports balls is now within reach.
\end{abstract}


\section{Introduction}
\label{S:intro}

The flow physics behind the phenomenon of drag reduction of dimpled spheres---golf balls---has been investigated since at least the 1970s \cite{bearman76,mehta85}.  Early studies relied primarily on wind tunnel experiments, typically collecting force data, and occasionally also performing flow visualizations with, for example, oil streaks. It is only recently, with the advent of large-scale Large Eddy Simulation (LES) and Direct Numerical Simulation (DNS) simulations of modest Reynolds numbers, that the accurate, predictive computational fluid dynamics (CFD) simulation of a golf ball has allowed deeper insight into the effects of dimples.  

The typical goal of a golf ball design is to maximize the range it can be driven in a straight line.  This primarily leads to the desire to reduce its drag as much as possible, with secondary goals of minimizing variation in side forces to maintain straight-line flight, and of maximizing the lift force produced by backspin.  Putting backspin on the ball produces lift via the Magnus effect, which extends the flight time and distance of the ball.  Drag reduction is mostly due to the dimples, which are sized to create a series of separation bubbles that will lead to early transition in the unstable shear layer above the bubbles.  The exact size, depth, and arrangement of dimples all contribute to the final aerodynamic properties of a golf ball under various conditions, and the full resolution of the flow details within each dimple are required to fully characterize these properties.

\section{Previous Studies}

The first noteworthy experimental investigation of the aerodynamics of golf balls under a variety of flow conditions is by Bearman and Harvey in 1976 \cite{bearman76}.  Their study used wind tunnel testing of scaled golf ball models to compare the characteristics of round vs. hexagonal dimples, with a smooth sphere used to assess the validity of their experimental setup.  The hexagonally dimpled ball also had far fewer dimples than the ``conventional" ball ($240$ vs. $330$ or $336$).  They found that the hexagonally dimpled ball had a lower drag coefficient ($C_D$) and higher lift coefficient ($C_L$) over most of the Re and spin rate range of interest, hypothesizing that the hexagonal dimples led to more discrete vortices due to the straight edges of the dimples.  The effect of the dimple edge radius was not studied.  For both dimple types however, they showed that the dimples serve to reduce the critical Re at which a drag reduction occurs, and that the drag coefficient remains nearly constant for a large range of Re after this point.  

Another detailed wind tunnel study was more recently performed by Choi et al. \cite{choi06}.  They studied both fully-dimpled and half-dimpled spheres without rotation.  In comparison to Bearman and Harvey, they used only round dimples with a much smaller depth ($k/d = 4 \cdot 10^{-3}$ for Choi et al. vs. $k/d = 9 \cdot 10^{-3}$ for Bearman and Harvey, where $k$ is the dimple depth and $d$ is the sphere diameter) and also with a larger number of dimples ($392$ vs. approximately $330$).  Their results showed a slightly higher critical $Re$ (${\sim}80\,000$ vs. ${\sim}50\,000$) with a slightly lower $C_D$ (${\sim}0.21$ vs. ${\sim}0.25$) afterwards, with a much more noticeable rise in $C_D$ after the initial drop.  Velocity data collected with a hot-wire anemometer was used to confirm that the turbulence generated by the free shear layer over the dimples led to an increase in momentum near the surface of the golf ball after reattachment, and that the separation angle remained at a constant $110^o$ after the critical $Re$.

Further studies have been performed using a combination of Reynolds--Averaged Navier--Stokes (RANS) \cite{ting02,ting03}, LES \cite{aoki10,li15,li17}, DNS \cite{smith10,beratlis12}, and wind tunnel experiments \cite{aoki10,chowdhury16}.  Li et al. proposed a link between small-scale vortices created at the golf ball dimples and a reduction in side-force variations at supercritical Reynolds numbers.  Also, several studies have discussed an apparent positive correlation between dimple depth the supercritical drag coefficient, and a negative correlation between dimple depth and critical Reynolds number \cite{ting03,chowdhury16,beratlis12}.  Although no specific mechanism has been proposed to explain the correlation with supercritical drag coefficient, the correlation with the critical Reynolds number is likely due to the instability of the free shear layer over the dimples becoming more unstable as the dimple (and hence separation bubble) becomes deeper, leader to quicker transition at lower Reynolds numbers. Numerous wind tunnel and computational experiments have also showed a strong positive correlation between lift force and spin rate (though smaller in magnitude than the negative lift force generated by a similar smooth sphere), and a slight correlation between drag and spin rate as well \cite{bearman76,aoki10,muto12,beratlis12}.

\section{Simulation Overview}

\subsection{Simulation Method}
\label{S:sim-method}

The flow of air around a golf ball is governed by the Navier--Stokes equations.  For the present study, we have utilized the implicit LES (ILES) method, sometimes referred to as under--resolved DNS.  The full viscous compressible Navier--Stokes equations are solved, with no additional models used to model the dissipation of kinetic energy at the smallest length scales present in the flow.  Instead, since the Reynolds number of the golf ball we simulate here --- $150\,000$ --- is low enough to capture most of the scales involved, we size the numerical grid in order to capture all but the smallest structures in the fluid flow, and allow the dissipation inherent to our spatial discretization method to `model' the dissipation that would occur at the smallest scales.  Such an approach is currently practical on modern computing clusters up to Reynolds numbers of approximately $500\,000$.  Furthermore, since we are going to the effort of capturing the vast majority of features in the flow, we use explicit time--stepping to capture the time evolution of these features; implicit methods, in addition to being memory intensive and complicated to implement for high--order schemes, would not greatly increase the maximum allowable time step given the small time scales that must be captured to preserve accuracy.

The spatial discretization method used here is the Flux Reconstruction (FR) approach, a unifying framework encompassing a variety of high-order methods, including the discontinuous Galerkin (DG) and spectral difference (SD) methods \cite{Huynh}.  The method uses Lagrange polynomials to represent the solution within each element of the grid.  However, in contrast to typical finite element methods, these polynomials are permitted to be \emph{discontinuous} between elements.  Within FR elements are coupled through the (approximate) solution of a Riemann problem.  This yields a common numerical flux at element interfaces which is then lifted into the interior of elements via specially defined  `correction functions'.  The method is covered in more detail in other works \cite{Vincent10, Vincent11, Castonguay12, David13, Asthana}, with an excellent summary being given in \cite{HandbookNumAnalysis}.

High order methods such as FR are particularly attractive for use in a DNS or ILES setting, and in fact have been shown to give remarkably good results in this context \cite{pypeta,pyfrstar,crabill18}.  Not only are they less dissipative, enabling the simulation of vortex-dominated flows with fewer degrees of freedom than a lower-order method, they are also far better suited for utilizing modern hardware than traditional second-order CFD methods.  This is due to the number of floating-point operations (FLOPs) performed per byte of memory accessed for each algorithm: an optimized second-order finite-volume solver will achieve less than 3\% of peak performance on Graphical Processing Units (GPUs) (which have an available FLOPs--to--bytes ratio of over 5) \cite{langguth13}, while a high--order discontinuous finite--element method (DFEM) is able to achieve over 50\% of peak performance on the same hardware \cite{pypeta}.

The novelty of our current approach lies in the use of these high--order methods on overset grids.  In the overset approach, a numerical grid is created around each body of interest, then the grids are patched together into a coherent hole through a process termed `domain connectivity'.  The key benefits of the approach are simplified mesh generation, and the ability to easily handle multiple objects in relative motion, such as the main rotor blades, tail rotor, and fuselage of a helicopter \cite{Helios,HeliosStrand}.  While traditional overset methods lose much accuracy at the boundaries between grids through the use of simple, low--order interpolation methods, the present approach maintains high--order accuracy across overset boundaries \cite{Galbraith13,crabill16,crabill18,CrabillThesis}, and is hence better at preserving complex, vortex--dominated flows such as those around rotorcraft, high--lift wing systems, and in the present case, spinning golf balls.

Additional consideration was given to efficient execution on modern, highly--parallel computing platforms such as NVIDIA graphical processing units (GPUs).  6 out of the top 10 fastest supercomputers in the world have either Xeon Phi or NVIDIA Tesla accelerators as of November 2017, and the use of accelerators is continuing to grow.  Since accelerators use different architectures than traditional CPUs, new approaches and algorithms are required to make efficient use of their available computing power.  To enable a degree of performance good enough to solve large cases on accelerators then, a new method for domain connectivity was developed and implemented in the overset connectivity library TIOGA \cite{tioga}, and combined with our in-house FR solver, ZEFR \cite{RomeroThesis}.  The performance of the combined system on static and moving grids has previously been shown to be high enough to solve large--scale flow physics problems on multi--grid systems in a reasonable amount of time \cite{crabill18,CrabillThesis}.

\subsection{Golf Ball Geometry}
\label{S:golf-geo}

\begin{figure}
\subfloat[Side View.]{
  \includegraphics[width=.29\textwidth]{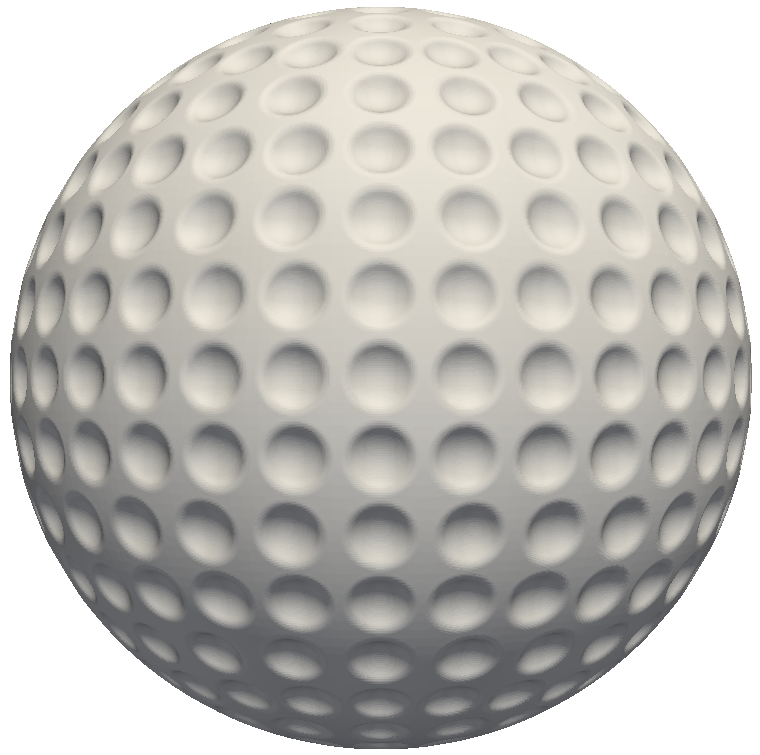}
  \label{fig:golf-side}
} \hspace{6pt}
\subfloat[Top View.]{
  \includegraphics[width=.29\textwidth]{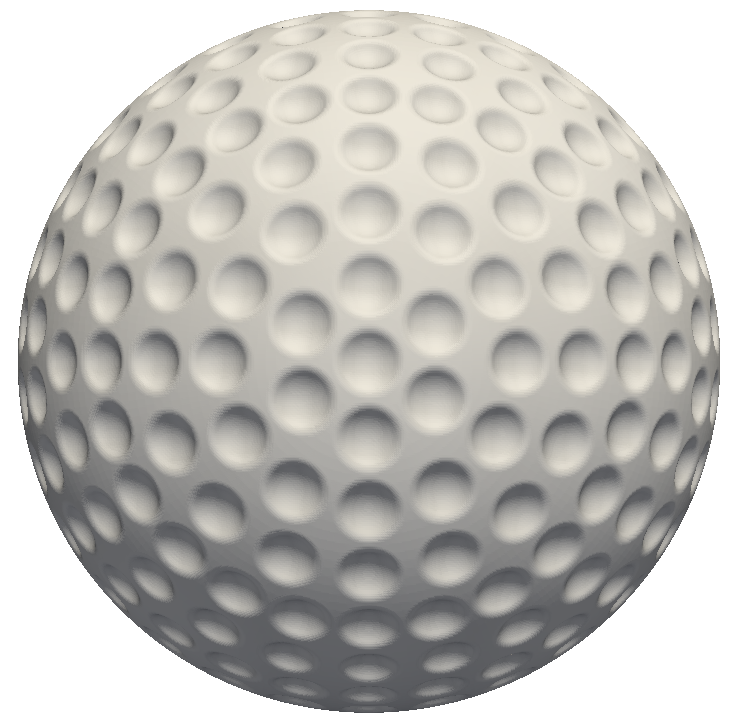}
  \label{fig:golf-top}
} \hspace{6pt}
\subfloat[Closeup of dimple mesh.]{
  \includegraphics[width=.29\textwidth]{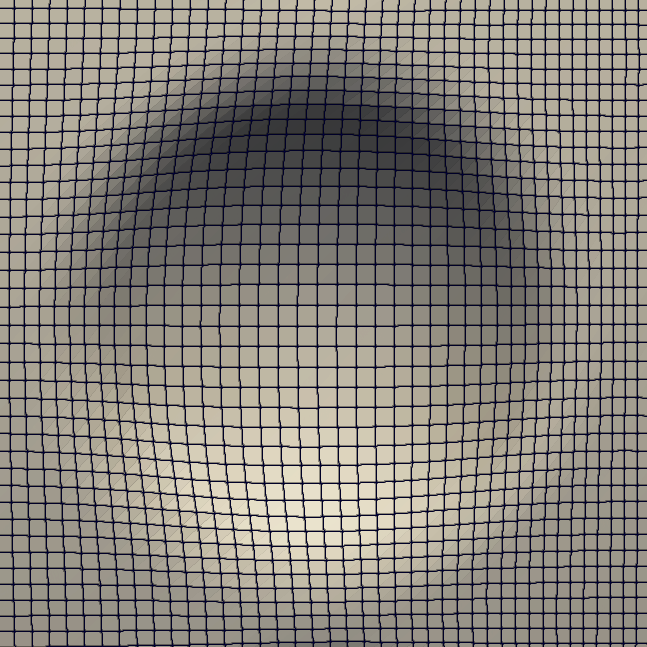}
  \label{fig:dimple-mesh}
} \\
\subfloat[Calculated $y^+$ values for near-wall solution points. Left, front, and right views.]{
  \includegraphics[height=.29\textwidth]{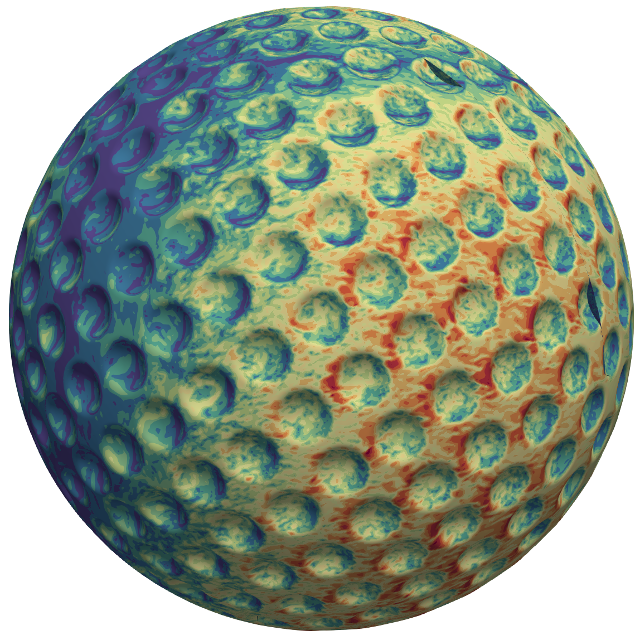}
  \hspace{6pt}
  \includegraphics[height=.29\textwidth]{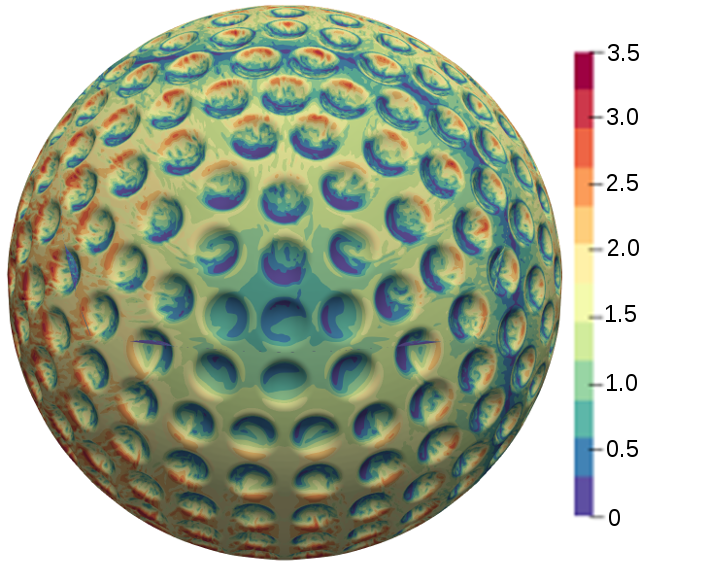}
  \includegraphics[height=.29\textwidth]{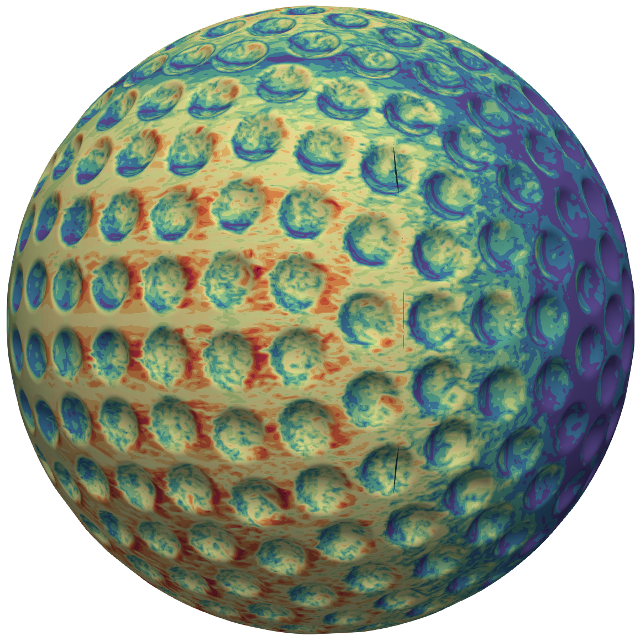}
  \label{sfig:yplus}
}
\caption{Golf ball surface and grid resolution.  Figure (c) shows the pre-agglomerated element sizes; the actual surface resolution of the golf ball is $4/3$ times greater than that shown.  The color mapping in (d) ranges from $y^+=0$ to $y^+=3.5$.}
\label{fig:golf-surf}
\end{figure}

In this study, the golf ball surface geometry was created as a parameterized CAD model with 19 rows of circular dimples (9 rows per hemisphere + 34 dimples around centerline), for a total of $388$ dimples, as shown in Figure \ref{fig:golf-surf}.  The golf ball diameter is $42.7$mm, the dimple depth is $6.41 \cdot 10^{-4}$m ($k/D = .015$), and the dimple diameter is a constant $2.99$mm ($c/D = 7.0 \cdot 10^{-2}$).  The dimple edges are filleted with a radius of $0.75$mm.  The surface was exported in the STL format and used within the multiblock structured mesh generator GridPro \cite{GridPro} to create a spherical grid with a boundary layer.  The surface of the golf ball was divided into $24$ roughly square regions, each with a resolution of $144 \times 144$ quadrilaterals, with $60$ layers in the radial direction, for a total of $29\,859\,840$ linear hexahedra, or $1\,105\,920$ cubically curved hexahedra after agglomeration.  The first cell height was chosen to be at an estimated $y^+$ value of 6.667 ($3.4 \cdot 10^{-5}$m), the first $18$ layers were held to a constant thickness, and the remaining $42$ layers were allowed to grow out to a final outer diameter of $31.82$mm.  The first cell height was chosen such that after agglomeration into cubically-curved hexahedra and run with 4th order tensor-product solution polynomials, the first solution point inside the element would lie at a $y^+$ of approximately 1. The surface mesh resolution was chosen to match the recommendations of Li et al. \cite{li15}, which are based upon recommendations from Muto et al. \cite{muto12}.  The golf ball grid of Li used a surface resolution of less than $\frac{1}{2} \delta_B$, where $\delta_B = 3 \sqrt{\frac{D \nu}{2 V}}$ is the estimated laminar boundary layer thickness $90^o$ from the stagnation point \cite{muto12} ($D$ is the diameter, $\nu$ is the kinematic viscosity, and $V$ is the freestream velocity).  Here, with $\delta_B \approx 2 \cdot 10^{-4} m$, our surface mesh resolution at the level of the linear grid is slightly more than $\frac{1}{2} \delta_B$, with the final resolution being slightly less than $\frac{1}{2} \delta_B$ once the high-order polynomials are introduced into the agglomerated hexahedra. Figure \ref{sfig:yplus} shows the actual distribution of $y^+$ for the solution points nearest the wall during the simulation, computed using the first cell height of $.12mm$ and $p=3$ Gauss--Legendre solution point locations.  Although the grid was originally designed to be run using 4th order instead of 3rd order polynomials, the maximum $y^+$ value is 3.5, and the average value over the entire surface is 1.

The mesh was output in the CGNS structured multiblock format and imported into HOPR (High-Order Pre-Processor) \cite{HOPR}, a utility which can agglomerate the cells of a structured mesh into high-order curved  hexahedra.  The new high-order mesh, in an HDF5-based HOPR-specific format, was then converted into the PyFR mesh format \cite{pyfr}, which ZEFR has the capability to read.

This pseudo-structured golf ball grid was then combined with a mostly Cartesian background grid created in Gmsh \cite{gmsh} to fill the desired extents of the full computational domain.  The box has a width and height of $0.6832$m (16 times the golf ball diameter $D$), and length $1.0248$m ($-12D$ to $12D$).  A refined region was created in the area to be occupied by the golf ball, with a refined wake region stretching out to the rear of the domain for a total of $715\,750$ linear hexahedra elements.

\section{Static Golf Ball}
\label{S:golf-static}

\begin{figure}
\centering
\includegraphics[width=.85\textwidth]{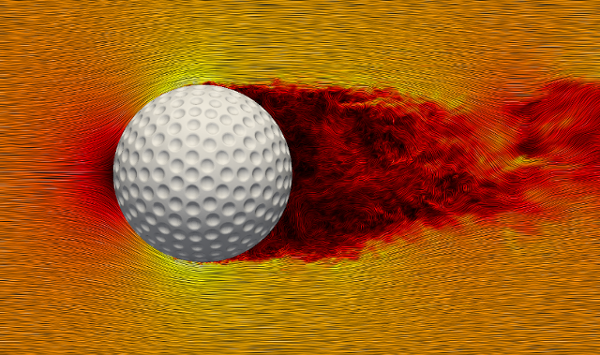}
\caption{View of approximate streamlines and velocity magnitude field through the $y=0$ plane (golf ball centerline).}
\label{fig:vel-static-1}
\end{figure}

\begin{figure}
\centering
\subfloat[Time history comparison]{
\includegraphics[width=.6\textwidth]{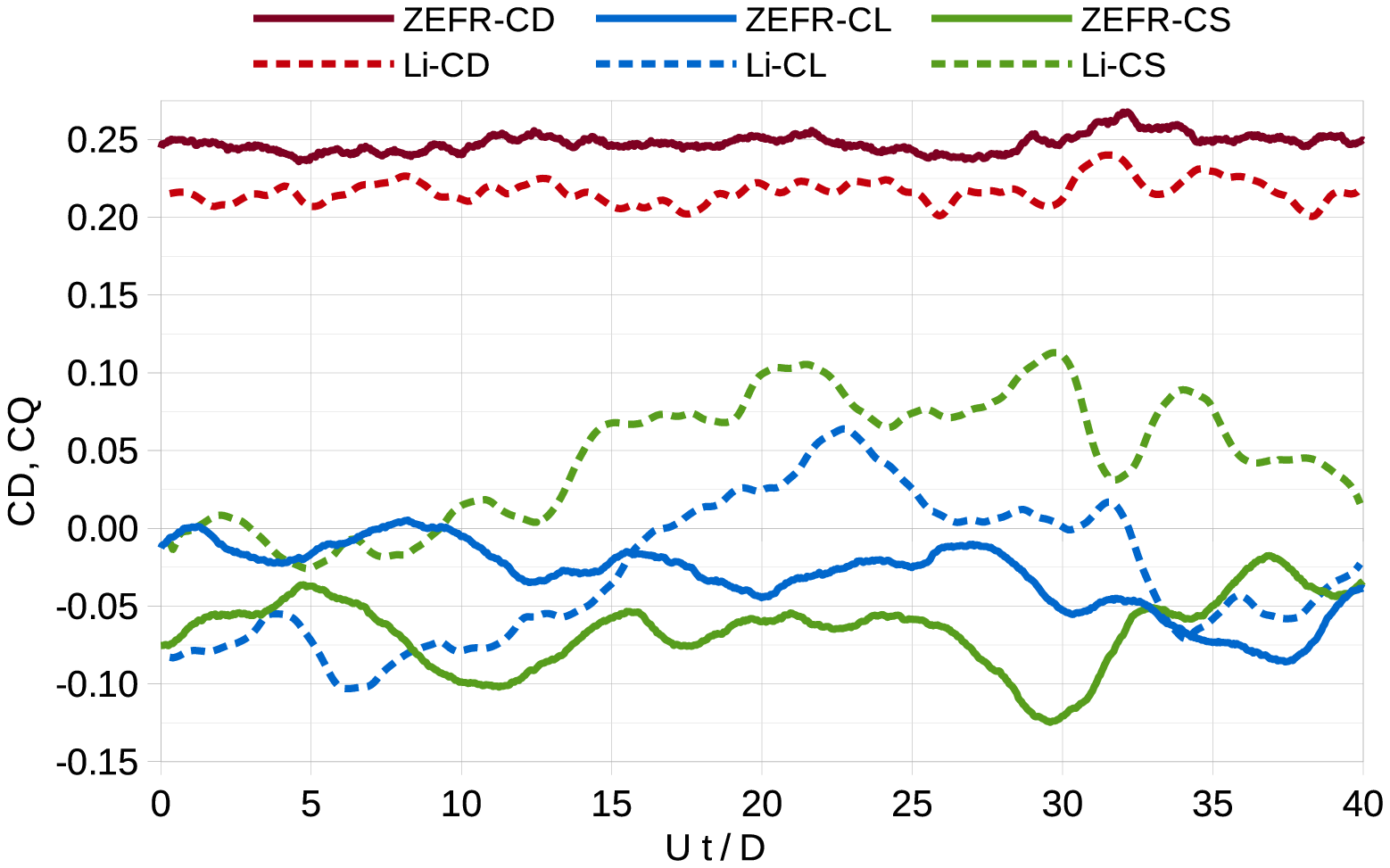}
\label{sfig:force-hist-compare}
}
\subfloat[Side forces polar comparison]{
\includegraphics[width=.35\textwidth]{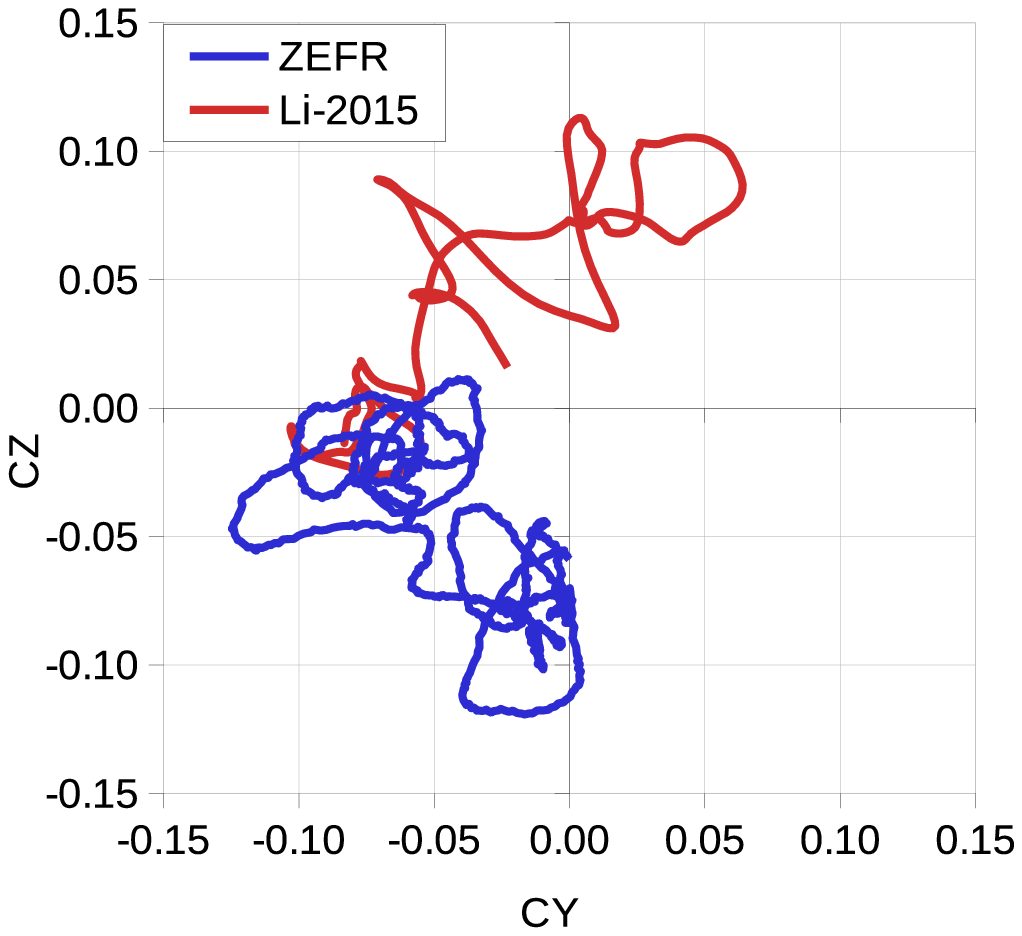}
\label{sfig:force-polar-compare}
}
\caption{Force coefficients for the static golf ball simulation, compared with those produced by Li et al. under similar conditions.}
\label{fig:force-compare}
\end{figure}

The simulation was advanced in time using the same adaptive RK54[2R+] scheme as before.  Third-order solution polynomials were utilized, as 4th order polynomials resulted in too restrictive of a time step on this grid to generate results in a reasonable amount of time.  The flow is along the $x$-axis, with a Reynolds number of $150\,000$ based upon the golf ball diameter of $0.0427$m, and a Mach number of $0.2$.  The full physical freestream conditions used (scaled such that the freestream velocity is 1) are shown in Table \ref{tab:golf-fs}.  An instantaneous view of velocity contours and approximate streamlines in the mid plane of the ball are shown in Figure \ref{fig:vel-static-1}.  The time histories of the drag and both side forces are shown in Figure \ref{sfig:force-hist-compare}, and a polar plot of the two side forces $C_Y$ and $C_Z$ are shown in Figure \ref{sfig:force-polar-compare}.

\begin{table}
\centering
\caption{\label{tab:golf-fs}Simulation conditions for all golf ball test cases.  The freestream quantities have been scaled such that no further non--dimensionalization is required for accuracy within the solver.}\vskip12pt
\begin{tabular}{r   l | r  l}
\toprule
Reynolds number &  $150\,000$    & $\rho$ &  $1.0$ kg/$\text{m}^3$ \\
Mach     &  $0.2$        & $V$    &  $1.0$ m/s \\ 
Prandtl  &  $0.72$       & $P$    &  $17.85714286$ Pa \\
$\gamma$ &  $1.4$       & $R$    &  $17.85714286$ J/(Kg K) \\
$L$      &  $0.0427$ m  & $T$    &  $1$ K \\ \bottomrule
\end{tabular}

\end{table}

As a verification that our results are correct, Figure \ref{fig:force-compare} also plots our force coefficient histories against those generated by Li et al. for a very similar case.  The conditions for their study were $Re = 110\,000$ incompressible flow; a lower Reynolds number than that used here, but still corresponding to the supercritical regime where the drag coefficient should remain nearly constant.  A second-order finite-volume LES solver was utilized for their simulation, using implicit time-stepping.  Their golf ball had $392$ dimples with a dimensionless diameter $c/D = 9.0 \cdot 10^{-2}$ and depth $k/D = 0.005$.  An unstructured prism / tetrahedron grid was used with a total of approximately $1.45 \cdot 10^6$ elements in the domain, with overall extents $-13D \leq x \leq 13D$ and $-5.6D \leq y,z \leq 5.6D$.  

Since the dimples primarily change the drag, it would be expected that the side forces should be quite similar between the two cases. Indeed, that is the case as shown in Figure \ref{sfig:force-hist-compare}; the average and standard deviation of the side force histories are nearly identical.  The side-force polar plot in Figure \ref{sfig:force-polar-compare} shows this as well; the two studies show similar trajectories, simply offset by a rotation about the axis of the flow. The present study was run for much longer ($100$ passes vs. $40$), leading to a more visibly bimodal polar plot, but the trends remain the same.  The drag histories are also in agreement; the offset between the two is to be expected, as the dimple depth used here is far greater than the dimple depth used by Li et al.  Results from a variety of studies have shown a direct correlation between dimple depth and supercritical drag coefficient, along with an inverse correlation to the critical Reynolds number.

\begin{figure}
\centering 
\subfloat[Through a row of dimples.]{
\includegraphics[width=.30\textwidth]{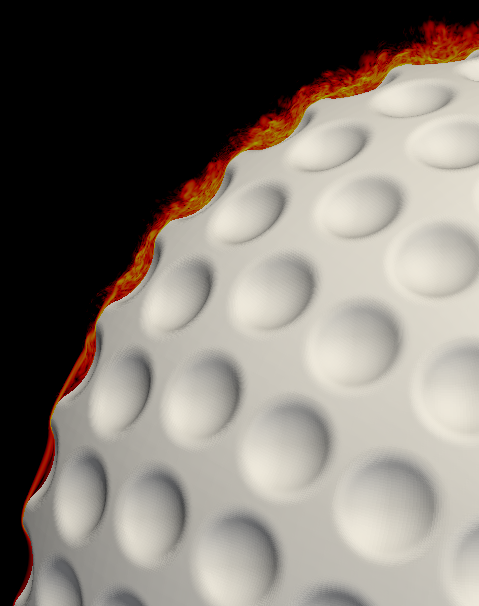}
} \hspace{6pt}
\subfloat[Between two rows of dimples.]{
  \includegraphics[width=.36\textwidth]{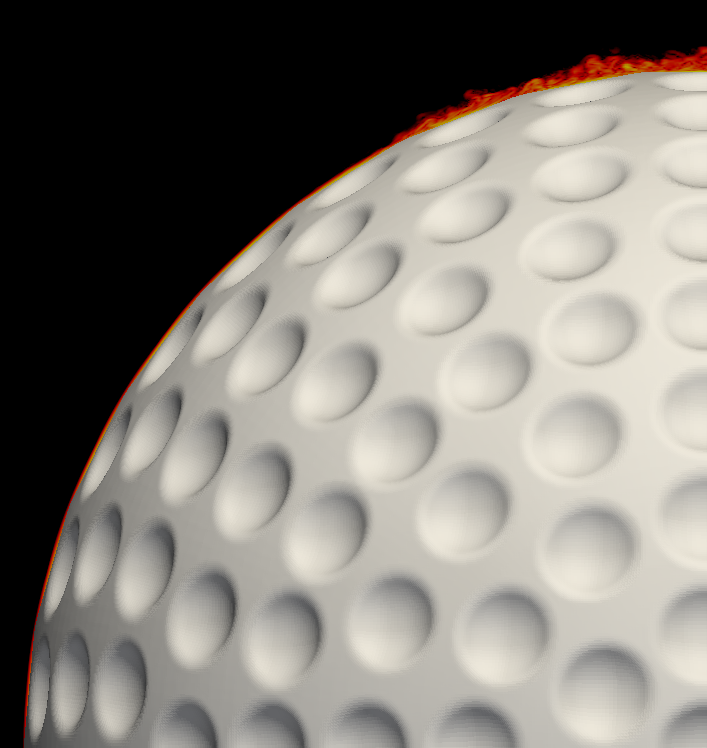}
  \label{sfig:cent-trans}
} \\
\subfloat[Closeup of slice through a row of dimples.]{
\includegraphics[width=.65\textwidth]{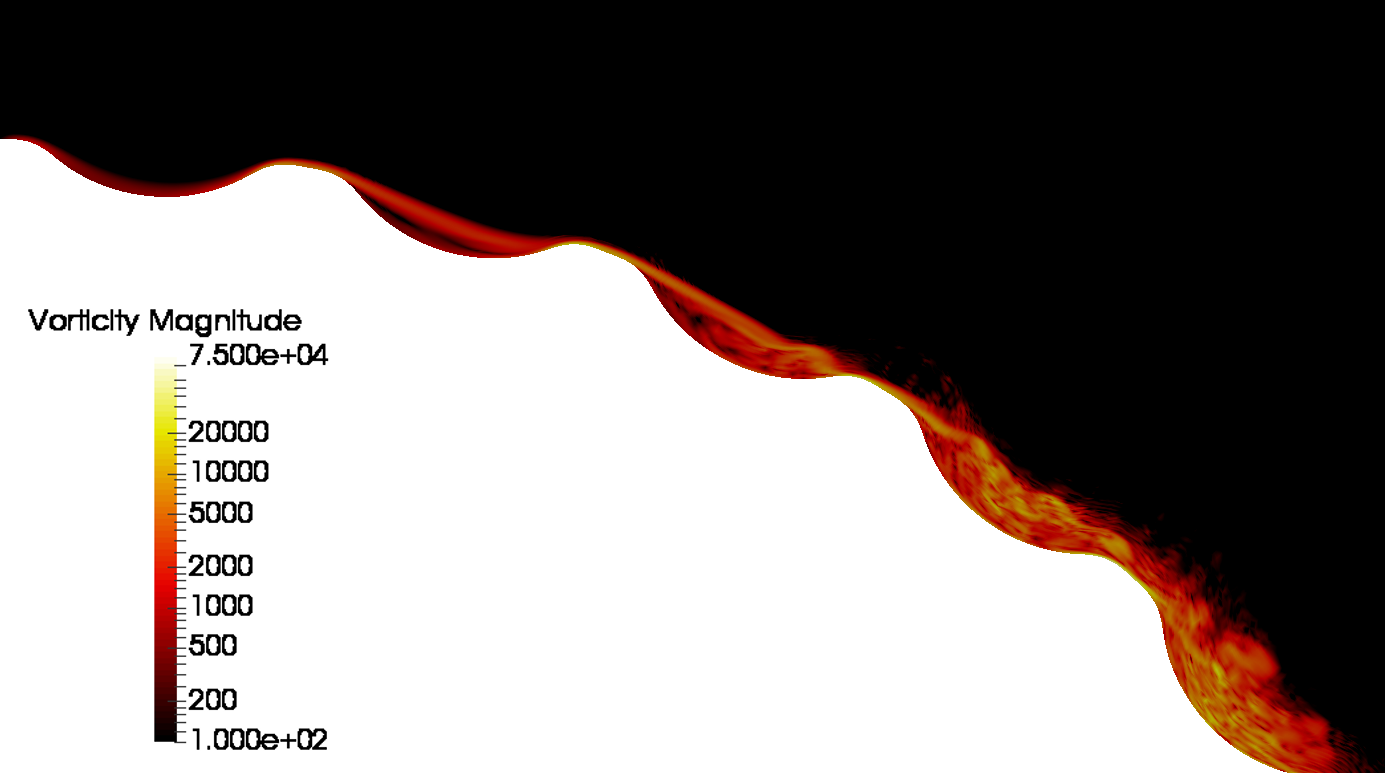}
}
\caption{Closeup view of log of vorticity magnitude through several slices near or at the $y=0$ plane, showing boundary layer transition occurring in the shear layer above a dimple.  In (b), the flow remains laminar until the point of transition clearly visible after the 6th dimple from the stagnation point.  In (c), the stagnation point is at the top left of the image.}
\label{fig:vort-static-1}
\end{figure}

\begin{figure}
\centering
\includegraphics[width=.4\textwidth]{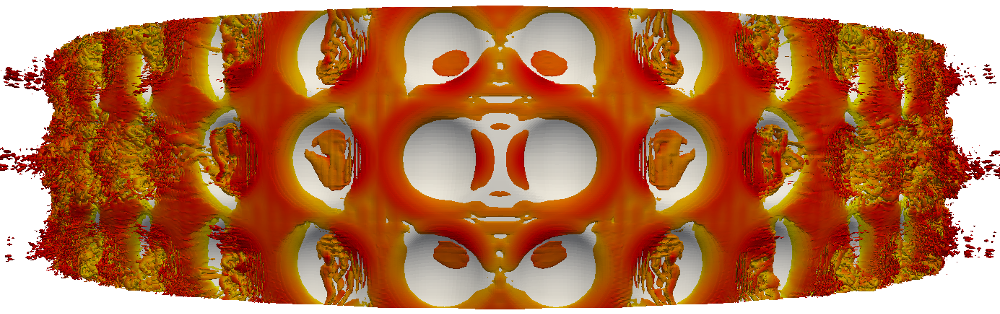} \\ \vspace{-2pt}
\subfloat[$z > 0$]{
\includegraphics[width=.4\textwidth]{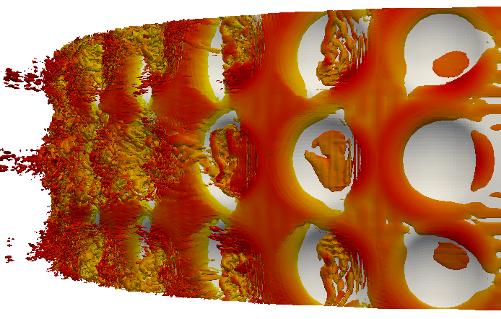}
\label{fig:q-vort-a}
} \hspace{6pt}
\subfloat[Closeup on transition areas. ($z<0$)]{
\includegraphics[width=.46\textwidth]{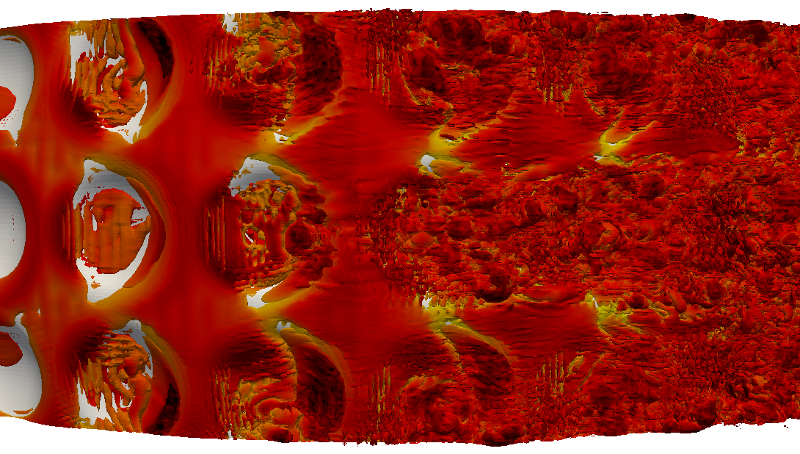}
\label{fig:q-vort-b}
}
\caption{Isosurfaces of Q-criterion colored by log of vorticity magnitude in a strip along the $y=0$ centerline of the golf ball.  Note that the location of transition is clearly visible as starting from the instabilities in the free shear layer over the dimples not far from the stagnation point.}
\label{fig:q-vort-1}
\end{figure}

In Figures \ref{fig:vort-static-1} and \ref{fig:q-vort-1}, we confirm the results of others in showing that the mechanism of transition is the growth of instabilities in the shear layer which forms over the recirculation regions inside the dimples not far from the stagnation point.  Figure \ref{fig:vort-static-1} shows contours of instantaneous vorticity magnitude through the $y=0$ plane (mid plane of the golf ball).  Separation and reattachment can be seen in dimples near the stagnation point, then near the top of the image, the shear layer breaks down and becomes turbulent.  Figure \ref{fig:q-vort-1} shows the same from a view above the centerline of the golf ball, with the three lines of dimples around $y=0$ shown looking down at the stagnation point.  Figure \ref{fig:q-vort-a} shows the strip for $z>0$ (the flow is roughly radially symmetric from the stagnation point).  The dimples near the stagnation point have well-defined reattachment areas; but when the flow reaches the next dimple, instabilities are visible in the shear layer over the dimple, clearly seen in Figure \ref{fig:q-vort-b}.  Over the 4th row of dimples (beginning at $31^\circ$ from the stagnation point), the shear layer has broken down and become fully turbulent.  The turbulent region then spreads out downstream from each dimple until the entire flow becomes turbulent by ${\sim}63^\circ$.  Figure \ref{sfig:cent-trans} shows the point at which the smooth line between two rows of dimples becomes turbulent.

\section{Spinning Golf Ball}
\label{S:golf-spin}

We next move on to the case of a spinning golf ball.  We keep the golf ball fixed at the origin, but apply a constant rotation rate around the $z$-axis; to fall in line with other studies, we choose a non-dimensional spin rate $\Gamma = \omega r / U_{inf} = 0.15$.  All other physical flow parameters are left the same.  In order to handle the movement of the golf ball grid, we use the arbitrary Lagrangian--Eulerian (ALE) form of the Navier--Stokes equations, which is as simple as adding an extra term to the convective fluxes due to the grid velocity.  We are here applying only a constant rotation rate around one axis, but future work may involve using our 6 degree of freedom capability to simulate the entire trajectory of the golf ball under the influence of its surface forces and moments.

\begin{figure}
\centering
\subfloat[]{\includegraphics[width=.55\textwidth]{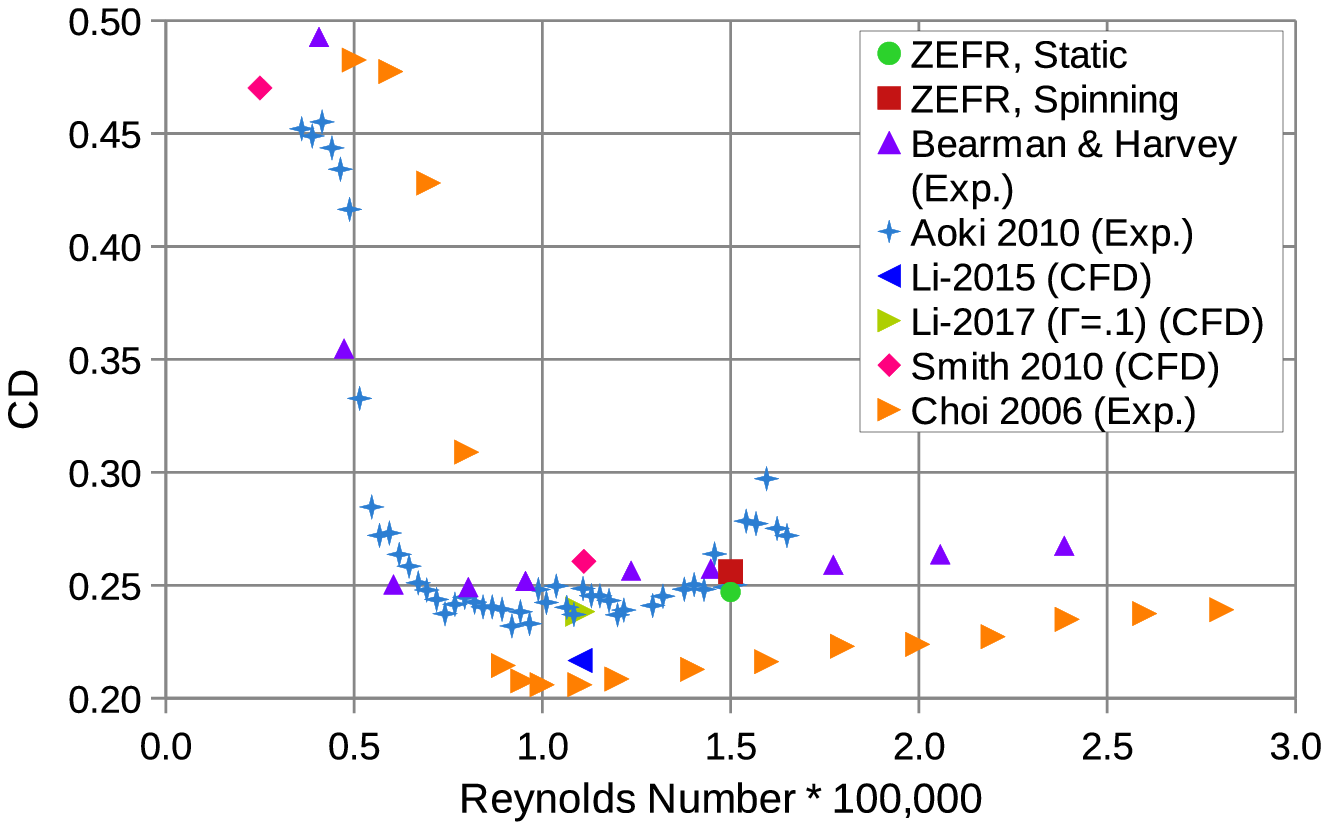}\label{sfig:CDplot}}
\subfloat[]{\includegraphics[width=.45\textwidth]{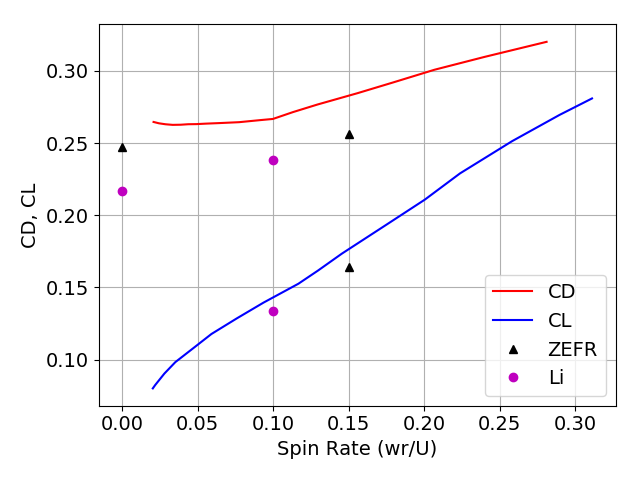}\label{sfig:CDCLplot}}
\caption{Comparison of $C_D$ and $C_L$ values from several experimental and computational studies alongside the present results produced with ZEFR.}
\label{fig:cd-study-compare}
\end{figure}

\begin{figure}
\centering
\subfloat[Time history.]{
\includegraphics[width=.6\textwidth]{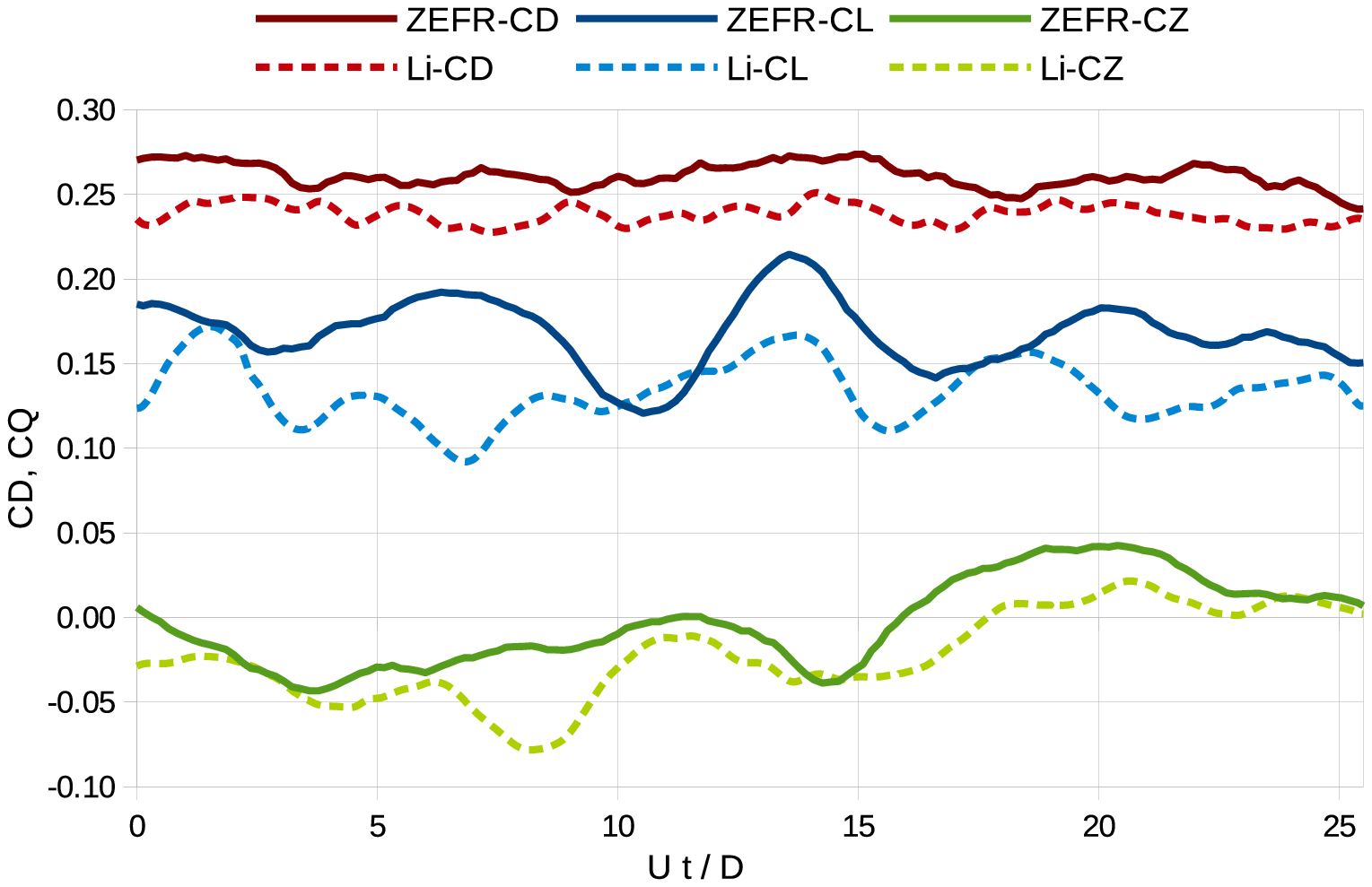}
\label{sfig:force-spin-hist-compare}
}
\subfloat[Side forces polar.]{
\includegraphics[width=.35\textwidth]{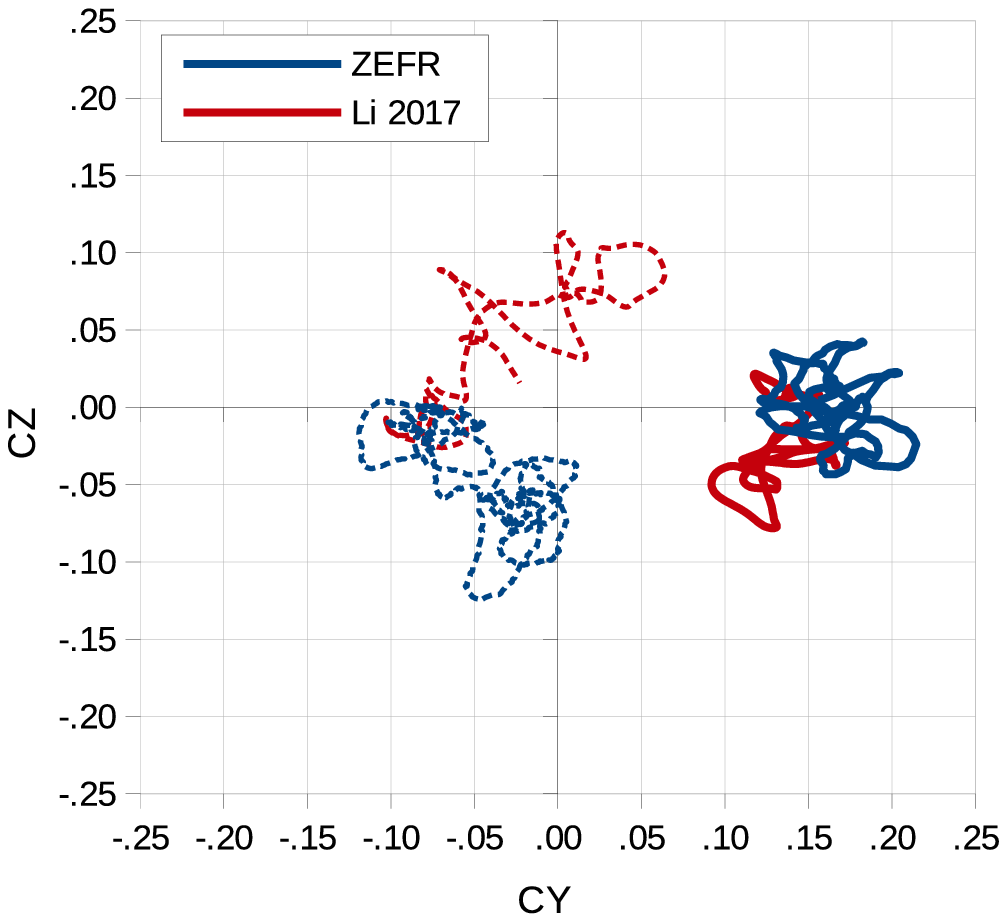}
\label{sfig:force-spin-polar-compare}
}
\caption{Comparison of force time histories for our spinning golf ball vs. the results from Li et al. 2017.  The dashed curves in (b) re-plot the same polar data from the static golf ball cases.}
\label{fig:force-spin-compare}
\end{figure}

Our average $C_D$ and $C_L$ values are compared against the results from a number of other studies, both experimental and computational, in Figure \ref{fig:cd-study-compare}.  As expected from previous literature, the spin induces a slightly higher drag coefficient than the static case but imparts a more regular variation in side forces upon the golf ball; the averages for all force coefficients (with standard deviations) are summarized in Table \ref{tab:forces} and the time history is shown in Figure \ref{fig:force-spin-compare}.  While the out-of-plane side force ($C_Z$) hovers near zero, the lift ($C_L$ or $C_Y$) hovers around a value of $0.16$, with relatively large low-frequency oscillations.  However, looking at a polar plot of the side forces, the oscillations are far more constrained than in the static case, where the symmetric nature of the flow allows the wake to oscillate randomly with no preferred direction.  In addition to providing a sizable lift force, the spin has the effect of imposing some structure and a more preferred direction to the oscillations of the wake.

As was done with the case of the static golf ball, we may compare against the prior results of Li et al. \cite{li17}, who have also performed detailed LES calculations of golf balls at a similar Reynolds number ($110\,000$ vs. $150\,000$) and spin rate ($.1$ vs. $.15$).  The time histories and polar plot of the force coefficient results from Li are plotted alongside the present results in Figure \ref{fig:force-spin-compare}.  The polar plots from the same cases without spinning are also included for comparison.  As in the static case, the present drag value is larger, as should be expected; our present value is about $5\%$ larger than that of the static golf ball.  The average lift value is also larger, as is also expected due to the higher spin rate used in the present study.  

The present values of lift and drag coefficients also agree well with those of Bearman and Harvery.  Using the data shown in Figure \ref{sfig:CDCLplot}, the estimated CD for a conventional golf ball at a nondimensional spin rate of .15 would be about .28, or $8\%$ higher than that of a static golf ball, with a lift coefficient of about .18.  Our results align more with their results at a spin rate of .13, with a lift coefficient of .16 and a spinning-to-static $C_D$ increase of $5\%$.  The results of Li et al, meanwhile, predict a much higher rise in $C_D$ with respect to spin rate, with a change of $11\%$ at a spin rate of .1.

\begin{figure}
\centering
\includegraphics[width=\textwidth]{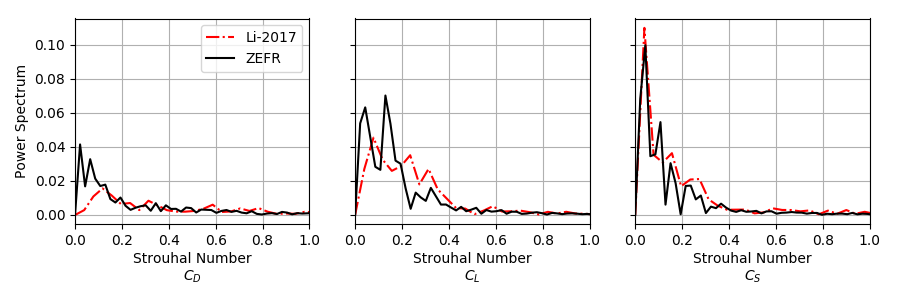}
\caption{Comparisons of power spectrum density for our spinning golf ball vs. the results from Li et al. 2017}
\label{fig:force-psd}
\end{figure}

We can also more quantitatively compare our results to those of Li by using the power spectrum of the golf ball forces, shown in Figure \ref{fig:force-psd}.  While the two side force spectra appear nearly identical, the lift and drag coefficients show slight differences.  In particular the drag spectra show a difference in peak, with the present results showing more low-frequency components.  Similarly, the present lift coefficient spectra also show a slight shift to lower frequencies for the two primary frequencies which appear, although the two smaller peaks are in the same locations as those of Li.

\begin{table}
\centering
\caption{Summary of average force coefficients for the static and spinning golf balls; $C_Q$ refers to the combined magnitude of the lift and side forces $C_Y$ and $C_Z$.  Present results compared to the similar study from Li et al.}\vskip12pt
\begin{tabular}{ccccc}
\toprule
 \;    & Static                        & Spinning ($\Gamma = .15$) & Li 2015 (Static) & Li 2017 ($\Gamma = .1$) \\ \midrule
$C_D$  & $\hphantom{-}0.2469 \pm 0.005$ & $0.256 \pm 0.010$ & $\hphantom{-}0.217 \pm 0.008$ & $\hphantom{-}0.238 \pm 0.0057$ \\ 
$C_Q$  & $\hphantom{-}0.076 \pm 0.020$  & $0.165 \pm 0.021$ & $\hphantom{-}0.079 \pm 0.019$ & $\hphantom{-}0.190 \pm 0.025$ \\ 
$C_Y$  & $-0.047 \pm 0.032$             & $0.164 \pm 0.021$ & $           -0.029 \pm 0.045$ & $\hphantom{-}0.134 \pm 0.018$ \\
$C_Z$  & $-0.044 \pm 0.032$             & $0.002 \pm 0.022$ & $\hphantom{-}0.046 \pm 0.040$ & $-0.022 \pm 0.026$ \\ \bottomrule
\end{tabular}
\label{tab:forces}
\end{table}

\begin{figure}
\centering
\includegraphics[width=.75\textwidth]{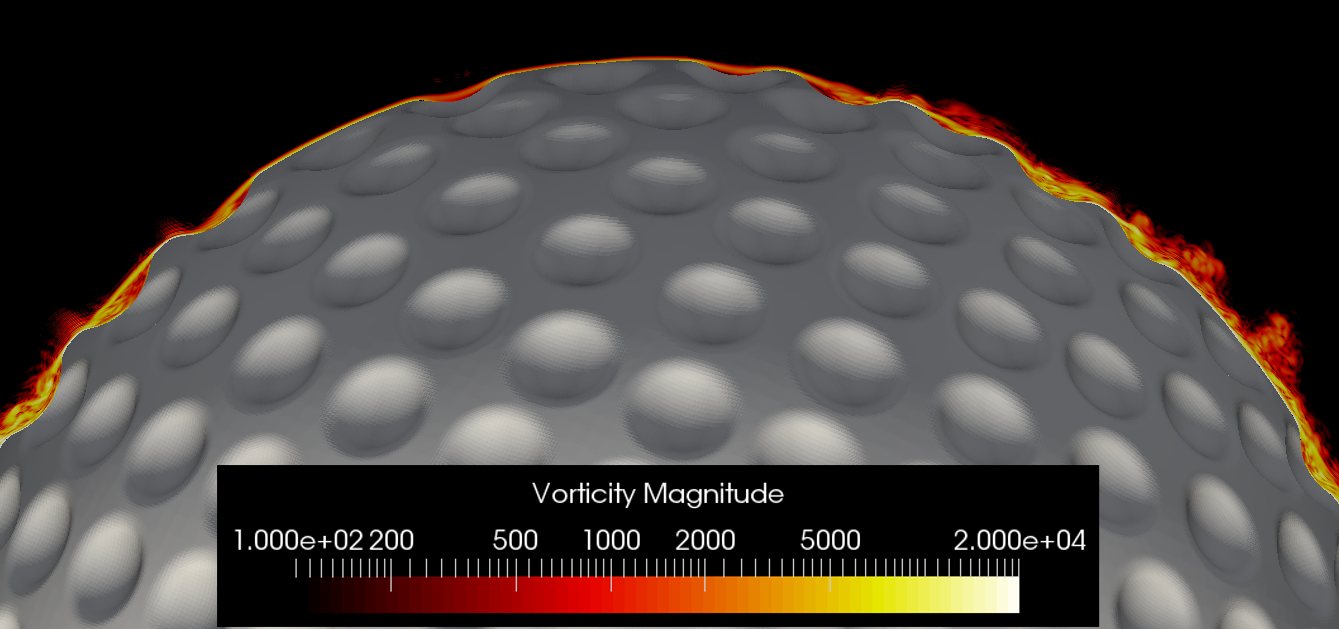}
\caption{Closeup view of $z=0$ centerplane showing log of vorticity magnitude.  The stagnation point is at the top of the image and the ball is spinning counter-clockwise.}
\label{fig:vort-trans-spin}
\end{figure}

\begin{figure}
\centering
\includegraphics[width=.8\textwidth]{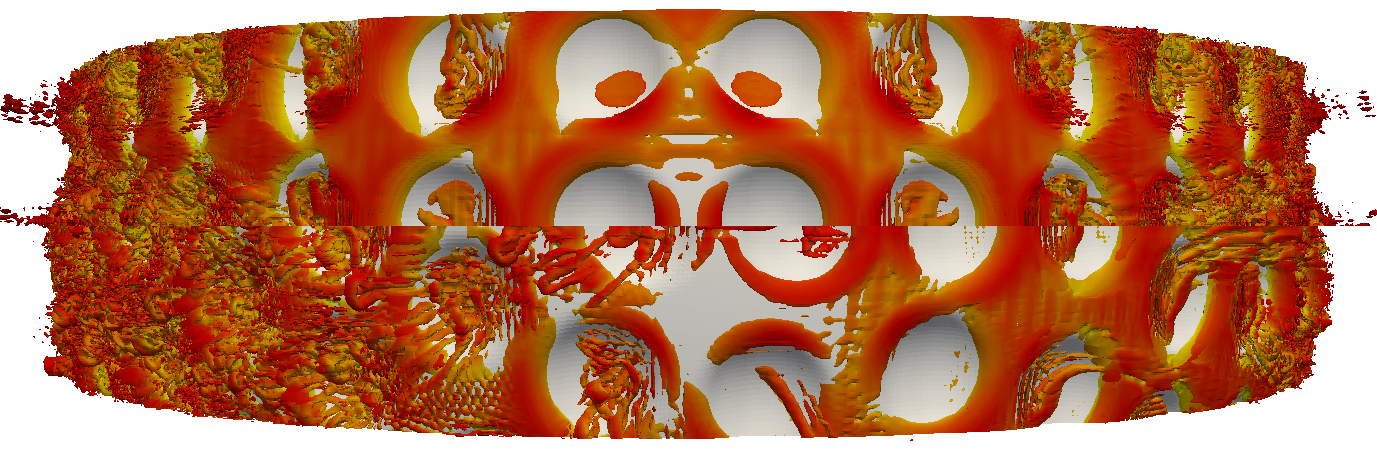}
\caption{Isosurfaces of Q-criterion colored by log of vorticity magnitude along the centerline of the golf ball. Top: Static golf ball, bottom: spinning golf ball at $T=49s$}
\label{fig:q-front-compare}
\end{figure}

While the overall flow features are quite similar from the static to the spinning golf ball, a few differences can clearly be seen.  Figure \ref{fig:vort-trans-spin} shows how the locations of transition shift; Figure \ref{fig:q-front-compare} likewise shows a direct comparison between flow features near the stagnation points of the static and spinning cases.  On the advancing side of the stagnation point, both figures show that the increased relative velocity over the surface of the golf ball lead to earlier transition; conversely, on the retreating side, the transition is slightly delayed relative to the static case.

\section{Conclusions}

In this work, the fluid dynamics of static and spinning golf balls were simulated using a high-order numerical scheme on overset grids. These are the first simulations of static and spinning golf balls using high--order numerical methods, and represent an advance in the state of the art in both scale-resolving CFD and overset grid calculations.  New algorithms were developed to leverage the modern hardware accelerators becoming increasingly common on large--scale computing clusters.  These new algorithms allow the  application of moving overset grids to high--order methods to solve large--scale fluid physics problems in a reasonable amount of time.  By comparing to a variety of previous experimental and computational studies, we have confidence that our present approach is able to accurately predict the complex, turbulent flow fields around golf balls and other sports balls, which operate at modest Reynolds numbers of less than $500\,000$.  Other sports applications are also within reach of high--fidelity simulation using our methods, such as hockey pucks, small or slow--speed sailboats, and bicycles or cyclists at modest speeds.  Beyond sports engineering, applications of interest to the aerospace community are also now within reach, including high-lift systems, turbomachinery, and a variety of multicopters and small-scale unmanned aerial vehicles.

\begin{acknowledgements}
The authors would like to acknowledge the Army Aviation Development Directorate (AMRDEC) for providing funding for this research under the oversight of Roger Strawn, the Air Force Office of Scientific Research for their support under grant FA9550-14-1-0186 under the oversight of Jean-Luc Cambier, and Margot Gerritsen for access to the XStream GPU computing cluster, which is supported by the National Science Foundation Major Research Instrumentation program (ACI-1429830).  We would also like to thank Dr. Peter Eiseman for providing academic licensing to the GridPro meshing software and assisting with the creation of several golf ball grids.  Lastly, we would like to thank Dr. Jay Sitaraman for his expertise and help on overset connectivity methods, and his help in ensuring our numerical methods were robust enough for broad applicability.
\end{acknowledgements}

\bibliographystyle{spmpsci}      
\bibliography{references}

\end{document}